\documentclass[aps,prd,10pt,preprint,superscriptaddress,onecolumn,nofootinbib]{revtex4-1}
\usepackage{graphicx, epsfig} 
\usepackage{amsmath,amssymb,amsfonts,dsfont,mathrsfs,amsthm,mathtools}
\usepackage{bm} 
\usepackage{color}
\usepackage[usenames]{xcolor}
\usepackage{hyperref}
\usepackage{siunitx}
\hypersetup{colorlinks=true,urlcolor=blue,linkcolor=magenta,citecolor=blue,filecolor=blue}
\usepackage[normalem]{ulem}
\usepackage{array}
\usepackage{hyperref}
\usepackage{booktabs}
\usepackage{blindtext}

\newcommand{\diff}[1]{\text{d}#1}

\newcommand{\Lag}{\mathscr{L}}

\begin{document}

\title{Higher-curvature generalization of Eguchi-Hanson spaces}

\author{Crist\'obal Corral}
\email{crcorral@unap.cl}
\affiliation{Instituto de Ciencias Exactas y Naturales, Facultad de Ciencias, Universidad Arturo Prat, Avenida Arturo Prat Chac\'on 2120, 1110939, Iquique, Chile}

\author{Daniel Flores-Alfonso}
\email{daniel.flores@cinvestav.mx}
\affiliation{Departamento de F\'isica, CINVESTAV-IPN, A.P. 14-740, C.P. 07000, Ciudad de M\'exico, Mexico}

\author{Gast\'on Giribet}
\email{gaston.giribet@nyu.edu}
\affiliation{Department of Physics, New York University, 726 Broadway, New York, NY10003, USA.}
\affiliation{Physics Department, University of Buenos Aires and IFIBA-CONICET, Ciudad Universitaria, Pabell\'on 1 (1428), Buenos Aires, Argentina}

\author{Julio Oliva}
\email{juoliva@udec.cl}
\affiliation{Departamento de F\'isica, Universidad de Concepci\'on, Casilla, 160-C, Concepci\'on, Chile.}

\begin{abstract}
We construct higher-dimensional generalizations of the Eguchi-Hanson gravitational instanton in the presence of higher-curvature deformations of general relativity. These spaces are solutions to Einstein gravity supplemented with the dimensional extension of the quadratic Chern-Gauss-Bonnet invariant in arbitrary even dimension $D=2m\geq 4$, and they are constructed out of non-trivial fibrations over $(2m-2)$-dimensional K\"ahler-Einstein manifolds. Different aspects of these solutions are analyzed; among them, the regularization of the on-shell Euclidean action by means of the addition of topological invariants. We also consider higher-curvature corrections to the gravity action that are cubic in the Riemann tensor and explicitly construct Eguchi-Hanson type solutions for such. 
\end{abstract}

\maketitle

\section{Introduction\label{sec:intro}}

Discovered in 1979 by Tohru Eguchi and Andrew Hanson~\cite{Eguchi:1978xp,Eguchi:1978gw}, and independently by Eugenio Calabi~\cite{Calabi:1979nuevo} the same year, the Eguchi-Hanson metric represents an interesting example of gravitational instanton~\cite{Hawking:1976jb, Eguchi:1976db, Gibbons:1978tef, Page:1978hdy, Hawking:1978ghb,Gibbons:1979xm, Eguchi:1979yx, Eguchi:1980jx}: It is a Ricci flat metric, and it corresponds to a non-compact, self-dual, Euclidean solution to Einstein equations in vacuum and with vanishing cosmological constant ($\Lambda = 0$); although an analog of the solution with non-vanishing $\Lambda $ also exists~\cite{Pedersen:1985}. When $\Lambda = 0$, the space is asymptotically locally Euclidean (ALE). More precisely, it is asymptotically $\mathbb{R}^4/\mathbb{Z}_2$. It admits a non-singular case whose topology is the cotangent bundle of the 2-sphere $\text{T}^*(S^2)=\text{T}^*(\mathbb{CP}^1)$, being $\mathbb{R} \times S^2$ at the so-called {\it origin} ($r=a$ in the coordinates \eqref{ansatz}-\eqref{fsolGR} below with $m=2$ and $\Lambda=0$) and asymptotically $\mathbb{RP}^3$ (at $r= \infty$). The projective space $\mathbb{RP}^3=S^3/\mathbb{Z}_2$ is isomorphic to the 3-dimensional rotation group $SO(3)\simeq SU(2)/\mathbb{Z}_2$, and the isometry group of the Eguchi-Hanson space is thus given by $  U(1) \times SU(2)/\mathbb{Z}_2 \simeq U(2)$. As it is for the $K3$ Calabi-Yau space, the reduced holonomy group of Eguchi-Hanson space is $SU(2)$, and this makes it possible to approximate the former space by combinations of copies of the latter. This is related to the fact that Eguchi-Hanson may be regarded as the resolution of the $A_1$ singularity, according to the ADE classification. In fact, Eguchi-Hanson has interesting applications in string theory in relation to the resolutions of orbifold singularities. It also has other interesting applications, such as serving as a {\it seed} for constructing higher-dimensional spaces with special holonomy~\cite{Higashijima:2001vk,Higashijima:2001fp,Higashijima:2002px,Clarkson:2005qx,Clarkson:2006zk,Giribet:2006xn,Wong:2011aa}. Multicenter, asymptotically $\mathbb{R}^4/\mathbb{Z}_k$ generalization of the Eguchi-Hanson instanton can also be constructed~\cite{Gibbons:1978tef}.

In this paper, we are interested in studying the higher-dimensional extension of the Eguchi-Hanson space, given by the metric~\eqref{ansatz}-\eqref{fsolGR} below. Considering the higher-dimensional generalization of Einstein spaces, however, makes it natural to consider, in addition to the Einstein tensor, higher-curvature contributions yielding second-order, covariantly conserved, rank-2 tensors in the field equations, cf.~\cite{Lovelock:1971yv}. Such contributions do not exist in 4 dimensions~\cite{Lanczos:1938sf} but are certainly possible in dimension 5 and higher~\cite{Lovelock:1972vz}. Here, we will take them into account. Higher-dimensional, higher-curvature gravitational instantons have been studied before; see for instance~\cite{Hendi:2008wq,Hendi:2012zg, Bueno:2018uoy, Corral:2019leh, Corral:2021xsu} and references therein and thereof. Here, we will explicitly construct generalizations of Eguchi-Hanson space in arbitrary even dimension $D$ and with quadratic curvature contributions. We will also consider examples with cubic curvature corrections.

The paper will be organized as follows: In Sec.~\ref{sec:EinsteinGravity}, we will study the higher-dimensional analog of the Eguchi-Hanson space with cosmological constant. In Sec.~\ref{sec:EGB}, we will introduce the higher-curvature terms we will consider. In Sec.~\ref{sec:6DEH}, we will present a six-dimensional version of the Eguchi-Hanson space, including the higher-curvature modification. We will discuss different version of the solution, exhibiting diverse topologies in the base manifold. In Sec.~\ref{sec:topologicalterms}, we will discuss the topological invariants and the regularized on-shell action associated to the six-dimensional solution. The $D$-dimensional solution will be presented in Sec.~\ref{sec:generalsolution}, where, in addition to the quadratic corrections; we will also consider cubic terms in the Riemann tensor in Sec.~\ref{sec:generalsolutionR3}.

\section{Higher-dimensional Eguchi-Hanson in Einstein gravity\label{sec:EinsteinGravity}}

For the sake of completeness, let us review first the $D$-dimensional Eguchi-Hanson solution in Einstein gravity. Inhomogeneous Einstein metrics were constructed as the non-trivial fibration over arbitrary K\"ahler-Einstein manifolds in Ref.~\cite{Hawking:1978ghb} (see also~\cite{Page:1985bq}) in even higher dimensions $D=2m$ with $m\in\mathbb{Z}$. A particular case is the higher-dimensional generalization of the Eguchi-Hanson metric \cite{Eguchi:1978gw}, whose line element is given by
\begin{align}\label{ansatz}
 \diff{s^2} = \frac{r^2}{4}f(r)\left(\diff{\tau} + \mathcal{B} \right)^2 + \frac{\diff{r^2}}{f(r)} + \frac{r^2}{4}\diff{\Sigma^2}\,,
\end{align}
where $\mathcal{B}=\mathcal{B}_\mu\diff{x^\mu}$ denotes the K\"ahler potential 1-form such that $\Omega=\diff{\mathcal{B}}$ defines the symplectic real form associated to the ($2m-2$)-dimensional K\"ahler base manifold with line element $\diff{\Sigma}^2$. In four dimensions, when the base manifold is $\mathbb{S}^2$, the metric~\eqref{ansatz} can be written in terms of the left-invariant Maurer-Cartan forms of $SU(2)$. In that case, the isometry group is $U(1) \times SU(2)/\mathbb{Z}_2$ and this solution is usually referred to as the gravitational analog of the Belavin-Polyakov-Schwartz-Tyupkin instanton in Yang-Mills theory~\cite{Belavin:1975fg}.

For concreteness, let us consider that the base space is the complex projective space $\mathbb{CP}^k$, with $k=m-1$. Then, the metric~\eqref{ansatz} is a solution of the $2m$-dimensional Einstein field equations with cosmological constant,
\begin{align}
    R_{\mu\nu} - \frac{1}{2}g_{\mu\nu}R + \Lambda g_{\mu\nu} = 0\,,
\end{align}
provided that the metric function $f(r)$ is given by
\begin{align}\label{fsolGR}
f(r) = \frac{2}{m}\left[1 - \left(\frac{a}{r}\right)^{2m} \right] - \frac{1}{2\left(m^2-1\right)}\,\Lambda r^2\,,
\end{align}
where $a$ is an integration constant. This metric is completely regular provided the parameters are chosen in such a way that there is no conical singularity at the bolt; the latter being defined as the  ($2m-2$)-dimensional set of fixed points located at the locus $f(r_b)=0$. Since the metric is positive definite in Euclidean signature, the range of the radial coordinate is $r\in \mathbb{R}_{\geq |r_b|}$. The regularity condition can be achieved by demanding periodicity on the $\tau $ coordinate, namely $\tau\sim\tau+\beta_\tau$, with $\beta_\tau$ defined through
\begin{align}\label{betasol}
 \beta_\tau = \frac{8\pi}{r_b\,f'|_{r=r_b}}\,,
\end{align}
where the prime denotes differentiation with respect to the radial coordinate $r$. 

In four dimensions ($m=2$) with a base manifold of $\mathbb{S}^2$ topology, the metric~\eqref{ansatz} is endowed with a Misner string~\cite{Misner:1963fr} along the $z$-axis as noticed in Ref.~\cite{Hawking:1998jf}. However, its position can be made unobservable by demanding extra conditions on the periodicity of $\tau$~\cite{Misner:1963fr}. For instance, the topology at $r=$ constant hypersurfaces becomes $\mathbb{S}^3/\mathbb{Z}_n$ if $0\leq\tau\leq4\pi/n$. Thus, the absence of conical singularities alongside the unobservability of the Misner string demands~\cite{Pedersen:1985}
\begin{align}\label{MS}
    \frac{8\pi}{r_b\,f'|_{r=r_b}} = \frac{4\pi}{n} \;\;\; \to \;\;\; r_b = \sqrt{-\frac{2(n-2)}{\Lambda}}\,.
\end{align}
The case with $\Lambda=0$ leads to the original Eguchi-Hanson metric where $\beta_\tau=2\pi$ and $r\geq a$, representing a space with $\mathbb{S}^3/\mathbb{Z}_2$ topology. However, depending on whether the cosmological constant is positive or negative, different topologies can be obtained. As shown in in Ref.~\cite{Chen:2020org}, the value of $r_b$ in Eq.~\eqref{MS} is modified in non-Einstein spaces with negative constant scalar curvature. Indeed, this is also the case in presence of higher-curvature corrections as we study here.

In four dimensions, this metric is globally (anti-)self dual in absence of the cosmological constant ($\Lambda = 0$). In that case, $\beta_\tau=2\pi$ and the action and topological invariants were obtained in Refs.~\cite{Eguchi:1978gw,Eguchi:1978xp}. Specifically,  its Eucliden on-shell action vanishes, its Euler characteristic is $\chi=2$, and its Hirzebruch signature is $\tau=-1$. This implies that there exists only one anti-self-dual $2$-form with the Atiyah-Patodi-Singer boundary conditions~\cite{Atiyah:1975jf,Atiyah:1976qjr,Atiyah:1976jg}. Moreover, since the index of the Dirac operator vanishes, there is no axial asymmetry whatsoever between left and right-handed Dirac spinors.  Nevertheless, there is an excess of two negative-chiralty spin $3/2$ spinors as it can be seen from the index of Rarita-Schinger operator~\cite{Eguchi:1978gw,Eguchi:1978xp}.

The Euclidean Taub-NUT metrics~\cite{Hawking:1976jb,Page:1978hdy} and the Eguchi-Hanson metric have similar off-diagonal components; they signal non-trivial circle bundles. Moreover, the Taub-bolt and Eguchi-Hanson spaces have the same type of degenerate submanifolds (bolts), although their global structures are different. Additionally, the self-dual Taub-NUT metric is hyper-K\"ahler, while, that of Eguchi-Hanson is K\"ahler.
In the presence of a negative cosmological constant, these two spaces differ in their asymptotic behavior: while the Taub-NUT metric is asymptotically locally hyperbolic, the Eguchi-Hanson metric is not~\cite{Chen:2020org}. These features represent the main differences between Taub-NUT and Eguchi-Hanson spaces.

In the following, we study how the higher-dimensional Eguchi-Hanson instanton is modified by the presence of higher-curvature corrections. In particular, we focus on terms of the Lovelock's series such that the field equations remain of second order.  

\section{Einstein-Gauss-Bonnet gravity\label{sec:EGB}}

On a $D$-dimensional manifold $\mathcal{M}$, the most general metric theory that is invariant under diffeomorphisms and local Lorentz transformations, which leads to second order field equations is known as the Lanczos-Lovelock theory of gravity~\cite{Lanczos:1938sf,Lovelock:1971yv}. Its dynamics is dictated by the action principle
\begin{align}\label{ILovelock}
 I[g_{\mu\nu}] = \sum_{p=0}^{\left[\frac{D-1}{2}\right]}\int_{\mathcal{M}}\diff{^Dx}\sqrt{|g|}\;\alpha_p\, \Lag^{(p)}\,,
\end{align}
where $g=\det g_{\mu\nu}$ is the metric determinant, $[\dots]$ represent the integer part, $\alpha_p$ are the Lovelock's couplings, and the $p$-th order Lagrangian is defined as
\begin{align}\label{Lovelockp}
 \Lag^{(p)} = \frac{1}{2^p}\,\delta^{\mu_1\ldots\mu_{2p}}_{\nu_1\ldots\nu_{2p}} R^{\nu_1\nu_2}_{\ \ \ \mu_1\mu_2}\dots R^{\nu_{2p-1}\nu_{2p}}_{\ \ \ \mu_{2p-1}\mu_{2p}}\,,
\end{align}
where $\delta^{\mu_1\ldots\mu_{k}}_{\nu_1\ldots\nu_{k}} = k!\,\delta_{[\nu_1}^{[\mu_1}\dots\delta^{\mu_k]}_{\nu_k]}$ denotes the completely antisymmetrized Kronecker delta of rank $k$. The field equations are obtained by performing stationary variations of the action~\eqref{ILovelock} with respect of the metric, giving
\begin{align}
 E^\mu_\nu \equiv -\sum_{p=0}^{\left[\frac{D-1}{2}\right]} \frac{\alpha_p}{2^{p+1}}\,\delta^{\mu\mu_1\ldots\mu_{2p}}_{\nu\nu_2\ldots\nu_{2p}}R^{\nu_1\nu_2}_{\ \ \ \mu_1\mu_2}\dots R^{\nu_{2p-1}\nu_{2p}}_{\ \ \  \mu_{2p-1}\mu_{2p}} = 0\,.
\end{align}
These equations of motion are at most of second order in derivatives of the metric and they propagate $D(D-3)/2$ degrees of freedom~\cite{Henneaux:1990au} around flat space; the same as general relativity in a $D$-dimensional space, the latter being the particular case when $\alpha_{p\geq 2} = 0$. 
Since we are interested in non-trivial fibrations of K\"ahler-Einstein manifolds, we shall focus on even-dimensional manifolds henceforth\footnote{Generalizations of Eguchi-Hanson spaces in odd dimensions were considered in \cite{Wong:2011aa}}. 

The so-called Einstein-Gauss-Bonnet theory can be obtained by truncating the Lovelock series to second order in $p$. Nevertheless, we will also keep the highest-order term in the Lovelock series \cite{Lovelock:1972vz} which
corresponds to the $2m$-dimensional Euler density \cite{Chern}; namely, $\Lag^{(m)}=\mathcal{E}_{2m}$ (see Eq.~\eqref{Eulertheorem} below). Being a topological invariant, the addition of the latter to the action does not modify the bulk dynamics, but it plays a crucial role in renormalizing the Euclidean on-shell action as well as in the computation of conserved charges in asymptotically locally AdS (AlAdS) spaces, providing a bulk counterterm once its corresponding coefficient is fixed~\cite{Aros:1999id,Aros:1999kt,Mora:2004kb,Kofinas:2006hr,Olea:2006vd,Arenas-Henriquez:2019rph}. Moreover, in four dimensions, when evaluated in configurations with (anti-)self-dual Weyl tensor, the topologically renormalized action yields an Euclidean on-shell action that is proportional to the Pontryagin index~\cite{Ciambelli:2020qny,Corral:2021xsu}, similarly as it happens for instantons in Yang-Mills theory. For non-conformally flat boundaries, additional counterterms based on the conformal completion of Einstein-AdS gravity may also be needed in order to render the Euclidean on-shell action and conserved charges finite~\cite{Anastasiou:2020zwc,Anastasiou:2020mik,Anastasiou:2021tlv}. Adding the Pontryagin density into the four-dimensional action with fixed coupling is also an interesting possibility, since it sets any the (anti-)self dual configurations as the ground state of the theory~\cite{Miskovic:2009bm,Araneda:2016iiy}.

Setting $\alpha_0 = -2\kappa\Lambda$, $\alpha_1=\kappa$, $\alpha_2=\kappa\alpha$, and $\kappa=(16\pi G_N)^{-1}$ with $G_N$ being the Newton's constant, the action becomes
\begin{align}\label{actionEGB}
 I_{\rm EGB}[g] = \kappa\int_{\mathcal{M}}\diff{^{2m} x}\sqrt{|g|}\,\left(R-2\Lambda + \alpha\,\mathcal{G} + \zeta\,\mathcal{E}_{2m} \right)\,,
\end{align}
where $\alpha$ and $\zeta$ are dimensionful coupling constants, and $\mathcal{G}$ is the Chern-Gauss-Bonnet term, namely,
\begin{align}\label{tongaEGB}
 \mathcal{G} = \frac{1}{4}\delta^{\mu\nu\lambda\rho}_{\alpha\beta\gamma\delta}R^{\alpha\beta}_{\  \mu\nu}R^{\gamma\delta}_{\  \lambda\rho} = R^2 - 4R^\mu_\nu R^\nu_\mu + R^{\mu\nu}_{\  \lambda\rho}R_{\  \mu\nu}^{\lambda\rho}\,.
\end{align}
In four dimensions, this term represents the bulk piece of the Euler characteristic [see Eq.~\eqref{Eulertheorem} below] and, therefore, it does not contribute to the bulk dynamics \cite{Lanczos:1938sf}. In dimensions higher than four, in contrast, this term does modify the field equations, leading to
\begin{align}\label{eomEGB}
R^\mu_\nu - \frac{1}{2}\delta^\mu_\nu R + \Lambda \delta^\mu_\nu + \alpha H^\mu_\nu = 0\,,
\end{align}
where $H^\mu_\nu$ is the contribution to the field equations of the Chern-Gauss-Bonnet term defined through
\begin{align}\label{HGB}
 H^\mu_\nu = -\frac{1}{8}\delta^{\mu\mu_1\ldots\mu_4}_{\nu\nu_1\ldots\nu_4}R^{\nu_1\nu_2}_{\ \ \mu_1\mu_2}R^{\nu_3\nu_4}_{\ \ \mu_1\mu_4} =  2R^{\mu\rho}_{\ \sigma\lambda}R_{ \ \nu\rho}^{\sigma\lambda}-4R_\rho^\sigma R^{\mu\rho}_{ \ \nu\sigma}+2RR^\mu_\nu - 4R^\mu_\lambda R^\lambda_\nu - \frac{1}{2}\delta^\mu_\nu \mathcal{G}\,.
\end{align}
Different analytic solutions to the field equations~\eqref{eomEGB} are known; these include black holes \cite{Boulware:1985wk, Cai:2001dz}, wormhole geometries \cite{Dotti:2006cp, Dotti:2007az}, Taub-NUT type gravitational instantons \cite{Hendi:2008wq} and many other examples.  In the next section, we provide a novel stationary solution to the field equations in Euclidean signature which represents a higher-dimensional generalization of the Eguchi-Hanson metric.

\section{Six-dimensional Eguchi-Hanson solution\label{sec:6DEH}}

Let us begin by considering the six-dimensional case, which is the simplest in which the quadratic curvature terms contribute in deforming the geometry in presence of an K\"ahler-Einstein base manifold. To solve the field equations~\eqref{eomEGB}, we focus on a six-dimensional Eguchi-Hanson-inspired ansatz~\eqref{ansatz}. In particular, we consider different topologies for the latter, these being fibrations over $\mathbb{CP}^2$, $\mathbb{CH}^2$, $\mathbb{T}^4$, $\mathbb{S}^2\times\mathbb{S}^2$, and $\mathbb{H}^2\times\mathbb{H}^2$.
More details of these geometries can be found in the next section. With these K\"ahler-Einstein base manifolds, the field equations~\eqref{eomEGB} turn out to be solved by the metric function
\begin{align}\label{solution}
 f_\pm(r) = \frac{2}{3}\left[ \gamma + \frac{r^2}{16\alpha}\left(1 \pm\sqrt{\left(1+3\Lambda\alpha \right) -\frac{128\sigma\alpha^2}{r^4}  + \frac{32\alpha a^6}{r^8}}\right)\right]\,,
\end{align}
where $a$ is an integration constant, $\gamma$ is related to the constant curvature of the base manifold, and $\sigma$ measures whether the latter is conformally flat ($\sigma=0$) or not ($\sigma=1$).\footnote{This is similar to what happens for solutions of the Einstein-Gauss-Bonnet field equations in the simpler geometries of the form $M_2\times_w\Sigma_{d-2}$, where $\times_w$ stands for a warped product. In that case, when $\Sigma_{d-2}$ is not conformally flat, there is an extra constant in the lapse function -- analogous to $\sigma $ here-- which leads to a slow decay of the metric at infinity relative to the $r$-dependent behavior that depends on the integration constant $a$, cf. Ref. \cite{Dotti:2005rc}.} Specifically, the values of these parameters can be summarized as follows
\begin{center}
\vbox{\begin{equation}\label{tabla1}
\begin{tabular}{cccccc}
\hline
 & $\;\;\;\mathbb{CP}^2\;\;\;$ & $\;\;\;\mathbb{CH}^2\;\;\;$ & $\;\;\;\;\;\mathbb{T}^4\;\;\;\;\;$ & $\;\mathbb{S}^2\times\mathbb{S}^2\;$ & $\;\mathbb{H}^2\times\mathbb{H}^2\;$ \\ 
 \hline
 $\;\;\;\gamma\;\;\;$ & $1$ & $-1$ & $0$ & $1$ & $-1$ \\ 
 \hline
 $\;\;\;\sigma\;\;\;$ & $0$ & $0$ & $0$ & $1$ & $1$ \\
 \hline
\end{tabular}
\end{equation}
}
\end{center}
The metric has a bolt at $r=r_b$ defined as the largest root of the polynomial $f(r_b)=0$ in Eq.~\eqref{solution}. Additionally, since the Euclidean signature implies that the metric is positive definite, the range of the radial coordinate is restricted to be $r\in \mathbb{R}_{\geq |r_b|}$. The absence of conical singularities is guaranteed as long as Eq.~\eqref{betasol} holds for the solution~\eqref{solution}. Asymptotically, the metric behaves as 
\begin{align}\notag
    f_{\pm }(r) &\simeq \frac{\left(1\pm\sqrt{1+3\Lambda\alpha}\right)r^2}{24\alpha} + \frac{2\gamma}{3} \mp \frac{8\sigma\alpha}{3\sqrt{1+3\Lambda\alpha}\;r^2}  \\ &\pm \frac{2\left[\left(1+3\Lambda\alpha\right)a^6-128\sigma^2\alpha^3\right]}{3\left(1+3\Lambda\alpha\right)^{3/2}\,r^6} + \mathcal{O}(r^{-10})\,.
\end{align}
Thus, one can directly notice that $\left(1\pm\sqrt{1+3\Lambda\alpha}\right)/(24\alpha)$ plays the role of effective cosmological constant, as long as $\alpha\neq0$. The solution is neither asymptotically conformally flat nor asymptotically locally AdS. In fact, the Weyl tensor squared behaves like 
\begin{align}
    W^{\mu\nu}_{\ \ \lambda\rho}W^{\lambda\rho}_{\ \ \mu\nu} \simeq \frac{2\left[2+3\Lambda\alpha\pm 2\sqrt{1+3\Lambda\alpha}\right]}{5\alpha^2} + \mathcal{O}\left(r^{-8} \right)\;\;\;\;\; \mbox{as} \;\;\;\;\; r\to\infty\,,
\end{align}
and it does not vanish at infinity. Notice that the branch $f_-(r)$ of the solution (\ref{solution}) is continuously connected to the solution of Einstein gravity [cf. Eq.~\eqref{fsolGR}] in the limit $\alpha\to0$, since
\begin{align}
 f_-(r) \simeq \frac{2}{3}\left(\gamma - \frac{a^6}{r^6}\right) - \frac{\Lambda r^2}{16} + \mathcal{O}(\alpha/r^2) \,.
\end{align}

Also notice that, on the curve of the parameter space defined by $\Lambda\alpha=-1/3$, the solution acquires a rather simple form; namely, 
\begin{align}\label{particularsolution}
 f_\pm(r) = \frac{2}{3}\left[ \gamma + \frac{r^2}{16\alpha}\left(1 \pm \frac{1}{3\Lambda r^2}\sqrt{ -\frac{96\Lambda a^6}{r^4} - 128\sigma}\right)\right]\,.
\end{align}
This is the confluent point where $f_+(r)=f_-(r)$ when $a=\sigma=0$ and, in many aspects, it exhibits certain features analogous to the so-called Chern-Simons point of Lovelock theory, cf. \cite{Zanelli:2005sa}. Nevertheless, it is worth emphasizing that this special curve on the parameter space is not the one that leads to a single maximally symmetric vacuum in the Einstein-Gauss-Bonnet theory; in the conventions are using, the maximally symmetric vacua coincide at $\Lambda\alpha=-5/12$ instead. For generic, fixed values of the K\"ahler potential $\mathcal{B}$, we are working with spaces that are not continuously connected with a maximally symmetric vacuum, regardless of the precise expression for the $f(r)$; nevertheless, there is still a particular value of the couplings for which the two branches coincide when both $\sigma$ and $a$ vanish. Since our spaces are not of constant curvature, the perturbations of the field equations around the solution at this special point may have support, in contrast to what happens with the perturbations around the maximally symmetric vacuum when both vacua coincide. 

On the curve of the parameter space where $\Lambda\alpha=-1/3$, the radial coordinate is bounded if $\sigma=1$ according to
\begin{align}\label{rbound}
    r_b\leq r \leq \left(-\frac{3a^6 \Lambda }{4} \right)^{\frac{1}{4}}\,,
\end{align}
with $\Lambda<0$ and $a\in\mathbb{R}$. When $\sigma=0$, the solution becomes
\begin{align}
f_\pm(r) = -\frac{\Lambda r^2}{8} + \frac{2\gamma}{3} \pm \frac{a^3}{r^2}\sqrt{-\frac{\Lambda}{6}}\,, 
\end{align}
and it exists only for $\Lambda<0$. The latter implies that both branches have a bolt at $f_\pm(r_b)=0$ depending on the values of the parameters $\gamma$ and $a$. In particular, for $\gamma=1$ and $\gamma=0$ we find that $f_+(r)$ and $f_-(r)$ have a bolt if $a<0$ and $a>0$,  respectively, and the space is thus completely regular. In contrast, if $\gamma=-1$ with $\sigma=0$ the metric exhibits a naked curvature singularity at $r=0$.

\section{Topological invariants and renormalized action\label{sec:topologicalterms}}

In this section, we compute the renormalized Euclidean on-shell action and Euler characteristic for the Eguchi-Hanson metric~\eqref{solution} with compact base manifolds $\mathbb{CP}^2$, $\mathbb{T}^4$, and $\mathbb{S}^2\times\mathbb{S}^2$. In the next section we will present the higher-dimensional generalization of the Eguchi-Hanson metric to arbitrary even dimension $D$, and, in particular, we will see that the solution in $D=8$ has also non-trivial Pontryagin index, as it happens in $D=4$.  

Let us first compute the Euler characteristic $\chi(\mathcal{M}_6)$ for the solution presented in Eq.~\eqref{solution}. In arbitrary even dimensions $D=2m$, the Euler theorem states that
\begin{align}\label{Eulertheorem}
   \int_{\mathcal{M}_{2m}}\diff{^{2m}x}\sqrt{|g|}\, \mathcal{E}_{2m} = \left( 4\pi\right)^m\,m!\,\chi\left(\mathcal{M}_{2m}\right) + \int_{\partial\mathcal{M}_{2m}}\diff{^{2m-1}x}\sqrt{|h|}\,\mathcal{C}_{2m-1}\,,
\end{align}
where $\mathcal{E}_{2m}\equiv\Lag^{(m)}$ [cf. Eq.~\eqref{Lovelockp}] is the Euler density in $D=2m$ dimensions, $\chi\left(\mathcal{M}_{2m}\right)$ is the Euler characteristic of the $2m$-dimensional manifold $\mathcal{M}_{2m}$, $h$ is the determinant of the induced metric $h_{\mu\nu}=g_{\mu\nu}-n_\mu n_\nu$ with $n^\mu$ being a space-like unit vector that is normal to the ($2m-1$)-dimensional boundary $\partial \mathcal{M}_{2m}$, and $\mathcal{C}_{2m-1}$ is the Chern form on $\partial \mathcal{M}_{2m}$ defined through the parametric integral as
\begin{align}\notag
    \mathcal{C}_{2m-1} &= 2m\int_0^1\diff{s}\,\delta^{\mu_1\ldots\mu_{2m-1}}_{\nu_1\ldots\nu_{2m-1}}K^{\nu_1}_{\mu_1}\left(\frac{1}{2}\mathcal{R}^{\nu_2\nu_3}_{\ \ \mu_2\mu_3} - s^2\,K^{\nu_2}_{\mu_2}K^{\nu_3}_{\mu_3} \right)\times \\
    &\quad \ldots\times\left(\frac{1}{2}\mathcal{R}^{\nu_{2m-2}\nu_{2m-1}}_{\ \ \mu_{2m-2}\mu_{2m-1}} - s^2\,K^{\nu_{2m-2}}_{\mu_{2m-1}}K^{\nu_{2m-1}}_{\mu_{2m-1}} \right)\,,
\end{align}
with $\mathcal{R}^{\mu\nu}_{\ \lambda\rho}$ and $K_{\mu\nu}=h^\lambda_\mu\nabla_\lambda n_\nu$ being the intrinsic and extrinsic curvature, respectively. These are related with the Riemannian curvature through the Gauss-Codazzi relation
\begin{align}
    \mathcal{R}^{\mu\nu}_{\ \lambda\rho} = h^\mu_\alpha h^\nu_\beta h^\gamma_\lambda h^\delta_\rho R^{\alpha\beta}_{\ \gamma\delta} + K^\mu_{\lambda} K^{\nu}_{\rho} - K^\mu_{\rho} K^{\nu}_{\lambda}  \,.
\end{align} 
Focusing on the compact cases of the solution~\eqref{solution}, a direct evaluation of the Euler characteristic in Eq.~\eqref{Eulertheorem} yields $\chi(\mathcal{M}_6)=3$, $\chi(\mathcal{M}_6)=0$, and $\chi(\mathcal{M}_6)=4$ for the fibered base manifolds $\mathbb{CP}^2$, $\mathbb{T}^4$, and $\mathbb{S}^2\times\mathbb{S}^2$, respectively. The latter can be equivalently obtained by means of the K\"unneth theorem, since $\chi(\mathbb{S}^2\times\mathbb{S}^2)=\chi(\mathbb{S}^2)\times\chi(\mathbb{S}^2) = 2\times2=4$.  

Here, we focus on the curve of the parameter space when $\Lambda\alpha=-1/3$ for the sake of simplicity. For conformally flat boundaries, the renormalized Euclidean on-shell action can be obtained by appropriately fixing the coupling constant $\zeta$ of the six-dimensional Euler density [cf. Eq.~\eqref{actionEGB}]. The latter plays the role of topological counterterms in even-dimensional AlAdS solutions. Although the Eguchi-Hanson solution in Einstein-Gauss-Bonnet gravity is not AlAdS, there is still a non-trivial value of the Euler density coupling that yields a finite result; this corresponds to $\zeta\Lambda^2=1/9$ in Eq.~\eqref{actionEGB}. On the other hand, when $\sigma=1$, the boundary is not conformally flat; nevertheless, the radial coordinate is naturally bounded according to~\eqref{rbound}, and so there is no need to include neither topological counterterms nor conformal completion to render the action finite. This means that in that case we can set $\zeta=0$ in a natural way. This is different to what happen when $\sigma=0$, for which the condition $\zeta\Lambda^2=1/9$ is needed for finiteness. Then, considering $\Lambda<0$ and the solutions with compact base manifolds, we obtain
\begin{align}
    I_{\rm EGB} &= \left\{\begin{matrix} 
    \frac{4\kappa a^9\beta_\tau\pi^2 }{3r_b^6}\sqrt{-\frac{6}{\Lambda}} & \mbox{for} & \mathbb{CP}^2 \,, \\
    \frac{2\kappa a^9\beta_{\tau}\beta_{\theta_1}\beta_{\phi_1}\beta_{\theta_2}\beta_{\phi_2}}{27r_b^6}\sqrt{-\frac{6}{\Lambda}} & \mbox{for} & \mathbb{T}^4 \,, \\
    -\frac{2\kappa\beta_\tau\pi^2 r_b^2\left[2 r_b^4 + \Lambda a^6 - \frac{r_b^4\Lambda}{12}\sqrt{-(8r_b^4+6\Lambda a^6)} \right]}{\sqrt{-(8r_b^4+6\Lambda a^6)}} & \mbox{for} & \mathbb{S}^2\times\mathbb{S}^2 \,.
    \end{matrix}
    \right.
\end{align}

The renormalized Euclidean on-shell action can be used in the path integral approach to study the contribution of these instantons, for instance resorting to the saddle point approximation~\cite{Gibbons:1976ue}. In the case of AlAdS spaces, the Euclidean path integral approach permits to explore the thermodynamics; for instance, it enables to study the well-known phase transitions that black holes develop in that sector at certain critical temperature~\cite{Hawking:1982dh}. A similar behavior may be found for solutions whose Euclidean version corresponds to AlAdS gravitational instantons~\cite{Johnson:2014xza,Johnson:2014pwa}. Thermodynamics and phase transitions of AlAdS solutions have also been studied in Lovelock theory, for instance in~\cite{Cvetic:2001bk,Nojiri:2002qn,Cai:2003gr,Clunan:2004tb,Dehghani:2005vh,Camanho:2013uda,Frassino:2014pha,Aranguiz:2015voa,Su:2019gby}. In the case of Eguchi-Hanson solutions, being intrinsically Euclidean and not AlAdS, the interpretation of the path integral results would be different. However, it would still be interesting to study whether any sort of phase transitions occurs in this sector. Instantons can contribute to the vacuum persistence amplitude in gauge theory~\cite{Osborn:1981yf}, and a similar computation can be performed in the case of gravitational instantons, cf.~\cite{Strominger:1984zy}. In the latter work, the authors provided a general formula for the one-loop determinant in terms of the renormalization group invariant masses, the volume of space, and topological invariants, by performing linear perturbations around the instantonic background. We believe these are interesting applications that can be studied using the analytic solutions found here.

In the next section, we generalize the Eguchi-Hanson instanton of Einstein-Gauss-Bonnet gravity to arbitrary even dimensions. Moreover, we show how the solution is modified in presence of cubic curvature terms in the Lovelock series.

\section{Generalization to higher dimensions\label{sec:generalsolution}}

Let us extend our solution to higher dimension. The general even-dimensional case, say $D=2m$ with $m\geq3$, can be solved analytically by considering different topologies of the base manifold. This was done in Sec.~\ref{sec:6DEH} for six dimensions as a particular case, and here we consider the metric Ansatz~\eqref{ansatz} with base manifold of topology $\mathbb{CP}^k$, $\mathbb{CH}^k$, $(\mathbb{T}^2)^{k}$, $(\mathbb{S}^2)^k$, and $(\mathbb{H}^2)^k$; the last three are the product of $k=m-1$ constant curvature 2-manifolds. Compact base manifolds with negative constant curvature can also be considered, for example, by taking spaces locally equivalent to $(\mathbb{H}^2)^k$ constructed out of quotients of the hyperbolic space by discrete subgroups. The K\"ahler potential and line element for each of these $(2m-2)$-dimensional base manifolds is given below. The field equations of Einstein-Gauss-Bonnet gravity~\eqref{eomEGB} are solved analytically by the line element~\eqref{ansatz} with the metric function
\begin{align}\label{laveintiseisb}
    f_\pm(r) &= \frac{2}{m}\Bigg\{\gamma + \frac{r^2}{16\alpha(m-2)}\left[1\pm\sqrt{1+8(m-2)\left(\frac{\Lambda\alpha m}{m^2-1}+\frac{4\alpha a^{2m}}{r^{2(m+1)}}-\frac{32\sigma\alpha^2(m-2)}{r^4(m-1)}\right)} \,\right] \Bigg\}\,.
\end{align}
Here $a$ is an integration constant, while $\gamma$ and $\sigma$ are parameters that depend on the topology and they admit the same geometric interpretation as the one given above Table~\ref{tabla1}. In particular, for the topologies considered in this Section, their values are
\begin{center}
\vbox{\begin{equation}
\begin{tabular}{cccccc}
\hline
 & $\;\;\;\mathbb{CP}^k\;\;\;$ & $\;\;\;\mathbb{CH}^k\;\;\;$ & $\;(\mathbb{T}^2)^{k}\;$ & $\;(\mathbb{S}^2)^k\;$ & $\;(\mathbb{H}^2)^k\;$ \\ 
 \hline
 $\;\;\;\gamma\;\;\;$ & $1$ & $-1$ & $0$ & $1$ & $-1$ \\ 
 \hline
 $\;\;\;\sigma\;\;\;$ & $0$ & $0$ & $0$ & $1$ & $1$ \\
 \hline
\end{tabular}
\end{equation}
}
\end{center}

K\"ahler spaces naturally allow for circle fiber bundles to be constructed on them. This is relevant for the construction of the higher-dimensional versions of the Eguchi-Hanson space discussed here.
K\"ahler manifolds are described by a Riemannian metric with associated line element $\diff{\Sigma}^2$ and a compatible symplectic closed 2-form $\Omega$. It is convenient to parameterize the symplectic form by a potential 1-form $\cal B$ defined by $\Omega=\diff{\cal B}$. For our purposes, we require the K\"ahler base manifolds to be, in addition, Einstein spaces. In two dimensions all spaces trivially fulfill these conditions; in higher dimensions, however, this is actually a stringent constraint. One way of constructing higher-dimensional K\"ahler-Einstein manifolds in arbitrary even dimension $2k$ is as the direct product of $k$ 2-dimensional spaces of equal curvature. For example, the product spaces $(\mathbb{T}^{2})^k$, $(\mathbb{S}^{2})^k$ and $(\mathbb{H}^{2})^k$ meet these requirements whenever their geometries are of the form detailed in the following Table: 
\begin{center}
\begin{tabular}{lllll}
\hline
 Space & \hskip2cm ${\cal B}_i$ \hskip2cm & \hskip2cm $\diff{\Sigma}_i^2$ \hskip2cm & \hskip1cm $\theta_i$ range \hskip1cm & \hskip1cm $\phi_i$ range \\ 
 \hline
 $(\mathbb{T}^2)^k$ & \hskip2cm $\theta_i\diff{\phi}_i$ \hskip2cm &  \hskip2cm $\diff{\theta}_i^2 + \diff{\phi}_i^2$ \hskip2cm & \hskip1cm $[0,\beta_{\theta_i}]$ \hskip1cm & \hskip1cm $[0,\beta_{\phi_i}]$ \\ 
 \hline
 $(\mathbb{S}^2)^k$ & \hskip2cm $\cos\theta_i\diff{\phi}_i$ \hskip2cm &  $\hskip2cm \diff{\theta}_i^2 +\sin^2\theta_i \diff{\phi}_i^2$ \hskip2cm & \hskip1cm $[0,\pi]$ \hskip1cm & \hskip1cm $[0,2\pi]$ \\
 \hline
 $(\mathbb{H}^2)^k$ & \hskip2cm $\cosh\theta_i\diff{\phi}_i$ \hskip2cm &  $\hskip2cm \diff{\theta}_i^2 +\sinh^2\theta_i \diff{\phi}_i^2$ \hskip2cm & \hskip1cm $(-\infty,\infty)$ \hskip1cm & \hskip1cm $[0,2\pi]$ \\
 \hline \\
\end{tabular}
\label{appendix:table1}
\end{center}

This Table shows a list of K\"ahler-Einstein spaces with symplectic potential ${\cal B}= \sum^k_{i=1}{\cal B}_i$ and a compatible Riemannian metric $\diff{\Sigma}^2= \sum^k_{i=1}\diff{\Sigma}^2_i$. For $k>1$, even-dimensional hyperspheres are Einstein spaces but not K\"ahler manifolds. In contrast, complex projective spaces $\mathbb{CP}^{k}$ are K\"ahler-Einstein manifolds for arbitrary $k$ when they are equipped with the Fubini-Study metric. Complex hyperbolic spaces with the Bergman metric also fulfill the geometric desiderata for our construction. Part of our calculations is benefited from considering the iterative construction of complex projective spaces~\cite{Hoxha:2000jf}. In this case, both the metric and the potential are written recursively; namely
\begin{subequations}\label{appendix:CPK}
\begin{align}
 \mathcal{B}_{(k)} &= (k+1)\sin^2\psi_k\left(\diff{\phi_k} + \frac{1}{k}\mathcal{B}_{(k-1)}\right),\label{appendix:Akpotential}\\
 \diff{\Sigma_{(k)}^2} &= 2(k+1)\bigg[\diff{\psi_k^2} + \sin^2\psi_k\cos^2\psi_k\left(\diff{\phi_k} + \frac{1}{k}\mathcal{B}_{(k-1)} \right)^2 
 + \frac{1}{2k}\sin^2\psi_k\diff{\Sigma_{(k-1)}^2} \bigg]\,, \label{appendix:CPkmetric}
 \end{align}
\end{subequations}
where $0\leq\psi_k\leq\pi/2$ and $0\leq\phi_k\leq 2\pi$. The base case ($k=1$) is provided by the sphere. The iterative formulae above take advantage of the fact that $\mathbb{CP}^k$ admits a foliation by $\mathbb{S}^{2k-1}$, modulo two points. The complex hyperbolic space $\mathbb{CH}^k$ admits a similar slicing; namely,
\begin{subequations}\label{appendix:CHK}
\begin{align}
 \mathcal{B} &= (k+1)\sinh^2\psi_k\left(\diff{\phi_k} + \frac{1}{k}\mathcal{B}_{(k-1)}\right),\label{appendix:Bkpotential}\\
 \diff{\Sigma^2} &= 2(k+1)\bigg[\diff{\psi_k^2} + \sinh^2\psi_k\cosh^2\psi_k\left(\diff{\phi_k} + \frac{1}{k}\mathcal{B}_{(k-1)} \right)^2 
 + \frac{1}{2k}\sinh^2\psi_k\diff{\Sigma_{(k-1)}^2} \bigg]\,, \label{appendix:CHkmetric}
 \end{align}
\end{subequations}
where $-\infty<\psi_k<\infty$ but all else is as in Eq. \eqref{appendix:CPK}.

As in the lower-dimensional examples, the bolt of the $2m$-dimensional solution we are constructing is defined as the ($2m-2$)-dimensional set of fixed points, located at the locus $f_\pm(r_b)=0$. This condition, alongside the absence of conical singularities, leads to a space that is geodesically complete. The branch $f_-(r)$ is continuously connected to the Einstein solution in arbitrary even dimensions [cf. Eq.~\eqref{fsolGR}] in the limit $\alpha\to0$. Additionally, defining $\Delta\equiv(m^2-1)\left[8\Lambda\alpha m (m-2)+m^2-1\right]$, we find that the asymptotic behavior of this metric is given by
\begin{align}\notag
    f(r) &\simeq \frac{\left[1\pm\sqrt{\frac{\Delta}{(m^2-1)^2}}\right]r^2}{8\alpha m (m-2)} + \frac{2\gamma}{m} \mp \frac{16\alpha\sigma m(m+1)}{(m-2)\sqrt{\Delta}\,r^2} \mp \frac{1024\alpha^3\sigma^2 m^3(m+1)^2}{(m-2)\sqrt{\Delta}\,r^6} + \mathcal{O}(r^{-10})\,,
\end{align}
as $r\to\infty$. The asymptotia of the branch $f_-(r)$ matches the Einstein gravity solution at large $r$ for any K\"ahler-Einstein base manifold. Solution 
(\ref{laveintiseisb}) is the generalized Eguchi-Hanson space in arbitrary even dimension $D=2m$, and in presence of quadratic curvature terms. 

\section{Generalizations with cubic terms and higher\label{sec:generalsolutionR3}}

The solution we have just presented can be extended to Lovelock theory with higher-curvature terms, cf. \cite{Wheeler:1985nh,Wheeler:1985qd}. As an example, in eight dimensions ($m=4$) consider a 6-torus base manifold $\mathbb{T}^6$ (i.e. $k=3$). For this ansatz, we can solve the equations of motion for the Lagrangian (\ref{Lovelockp}) with $p=0,1,2,3$ explicitly. That is to say, consider, in addition to the quadratic terms in (\ref{actionEGB})-(\ref{tongaEGB}), the following cubic term in the action
\begin{eqnarray}
I^{(3)}[g] = && \frac{\alpha_3\kappa}{8}\int_{\mathcal{M}}\diff{^{2m} x}\sqrt{|g|}\, \left(
R^3
-12RR_{\mu\nu}R^{\mu\nu}
+16R_{\mu\nu}R^{\mu}_{\,\, \rho}R^{\nu\rho }\right.
\nonumber \\
&&
\ \ \ \ \ \ \ \ \ \ \ \ +24R_{\mu\nu}R_{\rho\eta}R^{\mu\nu\rho\eta }+3RR_{\mu\nu\rho\eta}R^{\mu\nu\rho\eta}-24R_{\mu \nu }R^{\mu}_{\, \rho\eta\sigma }R^{\nu\rho\eta\sigma }
\label{dalearriba} \\
&&
\ \  \ \ \ \  \ \ \ \ \ \
\left.+4R_{\mu\nu\rho\eta} R^{\mu\nu\alpha\beta}R^{\rho\eta}_{\,\, \,\, \alpha \beta }
-8R_{\mu\nu\rho }^{\, \, \, \, \, \, \, \, \, \eta}R^{\mu\alpha \rho\beta }R^{\nu}_{\, \, \alpha \eta \beta}
 \right) \,.\nonumber
\end{eqnarray}
It can easily be checked that a polynomial equation implicitly determines the metric function $f(r)$. More precisely, in $D=8$, the field equations now take the form $G_{\mu\nu}+\Lambda g_{\mu\nu} + \alpha_2 H_{\mu\nu} + \alpha_3 M_{\mu\nu}=0$, where $H_{\mu\nu}$ is defined in Eq.~\eqref{HGB} and $M_{\mu\nu}$ comes from the variation of (\ref{dalearriba}); namely
\begin{align}
    M^\mu_\nu &= -\frac{1}{16}\delta^{\mu\lambda_2\ldots\lambda_7}_{\nu\rho_2\ldots\rho_7}R_{\  \ \lambda_2\lambda_3}^{\rho_2\rho_3}R_{\ \ \lambda_4\lambda_5}^{\rho_4\rho_5}R_{\ \ \lambda_6\lambda_7}^{\rho_6\rho_7}\,.
\end{align}
This yields the Wheeler type polynomial~\cite{Wheeler:1985nh,Wheeler:1985qd}
\begin{align}\label{elWheeeeler}
  384\alpha_3 r^4 f(r)^3 - 32\alpha_2 f(r)^2r^6 + r^8 f(r) + \frac{\Lambda}{30}r^{10} + a^8 = 0 \,,
\end{align}
where $a$ is an integration constant. To see how this polynomial compares to others like it in the literature, recall that the Eguchi-Hanson metric closely resembles that of Taub-NUT, and this similarity carries over to their respective generalizations. However, in spite of these parallels Eq. \eqref{elWheeeeler} is quite different from those found for Taub-NUT, see Ref.~\cite{Corral:2019leh}.

Furthermore, Eq. \eqref{elWheeeeler} is a polynomial equation that can be solved explicitly; however, its generic solution is cumbersome and not particularly illuminating. In order to have a clearer picture of the cubic solution, let us consider a particular case that enables us to visualize the form of the metric: Consider $\alpha_3={8\alpha_2^2}/{9}$, for which we find
\begin{align}
    f(r) = \frac{r^2}{32\alpha_2} - \frac{r^2}{32\alpha_2}\left(1+\frac{16\alpha_2\Lambda }{5}+\frac{96\alpha_2a^8}{r^{10}} \right)^{\frac{1}{3}}\,.\label{Lafifa}
\end{align}
This is an exact 8-dimensional solution of the cubic theory that behaves as the solution of $8$-dimensional Einstein gravity in the small $\alpha_2$ limit; namely $f(r)\simeq -\frac{\Lambda r^2}{30} - \frac{a^8}{r^8}+\mathcal{O}(\alpha _2)$. It is asymptotically equivalent to (\ref{ansatz}) with $\diff{\Sigma}^2$ being the flat metric on $\mathbb{T}^6$. Function (\ref{Lafifa}) has its root(s) at $r_b=(-30a^8/\Lambda )^{1/10}$. This is a concrete example of exact, analytic solution of the cubic Lovelock theory that asymptotically approaches a higher-dimensional generalization of Eguchi-Hanson space in presence of cosmological constant. Further examples can easily be obtained by following the method we described in this paper, i.e. by solving a polynomial of degree $n$, analogous to (\ref{elWheeeeler}), when the $2m$-dimensional extension of the $n$-dimensional Euler characteristics of order $\mathcal{O}(R^n)$, with $n\leq m-1$, are added to the gravity action.

\begin{acknowledgments}
We thank to Giorgos Anastasiou, Ignacio J.~Araya, Olivera Mi\v{s}kovi\'c, and Rodrigo Olea for insightful comments and discussions. The work of C. C. is partially supported by Agencia Nacional de Investigaci\'{o}n y Desarrollo (ANID) through FONDECYT grants No~11200025 and~1210500. D. F. is supported by a CONACYT postdoctoral fellowship. This work has been partially funded by CONACYT Grant No. A1-S-11548 and by Dirección de Postgrado (UdeC), through the grant UCO1866. G. G. is supported by CONICET and ANPCyT grants PIP-1109-2017, PICT-2019-00303. J.O. is partially supported by FONDECYT grants 1221504 and 1210635 and by Proyecto de cooperaci\'{o}n internacional 2019/13231-7
FAPESP/ANID.
\end{acknowledgments}

\bibliography{References}

\end{document}